\documentclass[journal=jctcce,manuscript=article]{achemso}
\usepackage{graphicx}
\usepackage{dcolumn}
\usepackage{xcolor}
\usepackage{adjustbox}
\usepackage{import}
\usepackage{booktabs}
\usepackage{dsfont}
\usepackage{mathtools}
\usepackage{comment}
\usepackage{amsmath}
\usepackage{multirow}
\usepackage{xcolor}
\usepackage{subcaption}
\usepackage{threeparttable}
\usepackage[section]{placeins}

\usepackage[table]{xcolor} 
\definecolor{headergray}{gray}{0.95}
\usepackage{lipsum}
\usepackage[colorlinks=true, linkcolor=blue, citecolor=blue, urlcolor=blue]{hyperref}
\usepackage{siunitx}   
\DeclarePairedDelimiter\bra{\langle}{\rvert}
\DeclarePairedDelimiter\ket{\lvert}{\rangle}
\DeclarePairedDelimiterX\braket[2]{\langle}{\rangle}{#1\,\delimsize\vert\,\mathopen{}#2}

\usepackage{bm}

\usepackage[utf8]{inputenc}
\usepackage[T1]{fontenc}
\usepackage{etoolbox}
\renewcommand*{\thesubsection}{\alph{subsection}}

\DeclareUnicodeCharacter{2500}{\textemdash} 
\DeclareUnicodeCharacter{2212}{-}

\usepackage{titlesec}
\titleformat{\section}[block]
  {\fontsize{14}{14}\bfseries}
  {\thesection}
  {1em}
  {}
\titleformat{\subsection}[hang]
  {\fontsize{14}{14}\bfseries}
  {\thesubsection}
  {1em}
  {}  

\author{Qasim Javed}
\affiliation{Department of Chemistry, University of California, Berkeley, California 94720, USA }

\author{Harrison Tuckman}
\affiliation{Department of Chemistry, University of California, Berkeley, California 94720, USA }

\author{Eric Neuscamman}
\email{eneuscamman@berkeley.edu}
\affiliation{Department of Chemistry, University of California, Berkeley, California 94720, USA }
\alsoaffiliation{Chemical Sciences Division, Lawrence Berkeley National Laboratory, Berkeley, CA, 94720, USA}

\title
  {Aufbau Suppressed Coupled Cluster Theory for Doubly Excited States}

\begin{document}








\begin{abstract}
We generalize the Aufbau suppressed coupled cluster formalism into the realm of doubly excited states by deriving, implementing, and testing a wave function initialization strategy that allows the zeroth order wave function to match the largest configurations of a doubly excited reference wave function while maintaining the method's overall asymptotic cost parity with ground state singles and doubles theory.  Starting from state-averaged complete active space self consistent field references, this approach produces highly accurate excitation energies for states dominated by a single doubly excited determinant, as well as states in glyoxal and similar molecules where two different doubly excited determinants have large weights.  Typical excitation energy errors in both types of states are on the order of 0.15 eV, with the largest observed error being 0.3 eV.  These errors stand in stark contrast to equation of motion methods, where typical errors are 4 to 6 eV at the singles and doubles level and 0.4 to 0.8 eV at the full triples level.  It remains an open question how best to generalize the Aufbau suppression approach into an even wider variety of multi-configurational double excitations, but these early results offer strong motivation for further investigation.
\end{abstract}

\maketitle

\section{\label{sec:level1}Introduction }


Electronic states with significant double excitation character play important
roles in spectroscopy and photochemistry, for example in singlet fission,
\cite{smithSingletFission2010,https://doi.org/10.1002/cptc.202000211}
thermally activated delayed fluorescence,
\cite{desilvaInvertedSingletTriplet2019,aizawa2022delayed}
and the post-excitation relaxations of carotenoids.
\cite{polivka2004ultrafast}
Often optically dark in experiments due to symmetry,
\cite{oliver2015following}
double excitations are an important example of where
theoretical modeling has the opportunity to play an especially helpful role.
However, these states have long resisted easy modeling with the most widely
used quantum chemistry methods.
Time-dependent density functional theory
\cite{doi:10.1142/9789812830586_0005,PhysRevLett.52.997,PhysRevLett.76.1212,10.1063/1.1904586,annurev:/content/journals/10.1146/annurev-physchem-032511-143803,doi:https://doi.org/10.1002/9781119417774.ch2,10.1093/acprof:oso/9780199563029.001.0001}
is unable to see them at all within its standard adiabatic approximation,
\cite{annurev:/content/journals/10.1146/annurev-physchem-082720-124933,10.1063/1.1651060}
while coupled cluster (CC) theory's linear-response (LR)
\cite{koch1990coupled,koch1990excitation,pedersen1997coupled}
and equation of motion (EOM)
\cite{10.1063/1.452039, https://doi.org/10.1002/wcms.76,10.1063/1.464746,RevModPhys.79.291,Shavitt_Bartlett_2009,doi:https://doi.org/10.1002/9781119417774.ch4}
paradigms typically make multi-eV errors in doubly excited excitation energies, at
least when working at the relatively affordable singles and doubles level.
\cite{kossoskiReferenceEnergiesDouble2024,10.1063/1.3236843,10.1063/1.1378323,doi:10.1021/acs.jctc.8b01053,doi:10.1021/ct300591z,doi:10.1021/jacs.7b06283}
These difficulties are essentially both products of the
fact that promoting two rather than just one electron puts a state
farther from the ground state, at least in the ways that matter
for LR theories.
State-specific CC methods, in contrast, do not face the same
constraints, and can achieve high accuracies
even at the singles and doubles level.
\cite{10.1063/1.5128795}
In this work, we explore whether a recently-introduced state-specific CC
framework, Aufbau suppressed coupled cluster (ASCC),
\cite{tuckmanAufbauSuppressedCoupledCluster2024,tuckmanImprovingAufbauSuppressedCoupled2025}
is able to succeed in the doubly excited context.
We begin by testing how well double excitations dominated by a single
determinant can be encoded into the ASCC ansatz,
after which we also perform preliminary tests in
a handful of simple but nontrivial
multi-configurational cases.
We find that, in both types of states, singles and
doubles ASCC is capable of delivering doubly excited state accuracies on
par with what CC methods typically deliver for singly excited states.


The difficulties in modeling doubly excited states with many popular excited
state CC methods are well established.
In the CC2 approach, \cite{CHRISTIANSEN1995409} the doubles-doubles
block of the CC response Jacobian ends up approximated as a diagonal matrix
containing ground state orbital energy differences,
\cite{winter2011scaled}
preventing  accurate descriptions of doubly excited states' correlation and
orbital relaxation effects.
In the singles and doubles EOM approach (EOM-CCSD),
the doubles-doubles block is in better shape, but the doubles-triples
block is missing, preventing the theory from capturing the key
post-excitation relaxation effects that its doubles operator
provides for singly excited states. \cite{10.1063/5.0091715}
Adding triples (EOM-CCSDT) addresses this issue, but, somewhat surprisingly
at least from a formal perspective, often fails to bring excitation
energy errors below 0.5 eV. \cite{kossoskiReferenceEnergiesDouble2024}
In contrast, by treating doubly excited states in a state-specific manner,
the $\Delta$CC approach is able to do significantly better,
at least for states dominated by a single determinant.
\cite{10.1063/1.5128795,doi:10.1021/acs.jctc.4c00034}
However, $\Delta$CC's single-configurational nature makes it difficult
to apply to states in which multiple configuration state functions (CSFs)
have large weights.
Similarly,
the quasi-state-specific intermediate state EOM approach that builds atop two-determinant coupled cluster (TDCC)
has shown excellent accuracy for one-CSF double excitations,
\cite{10.1063/5.0091715}
but its early results on multi-CSF states (such as the glyoxal state
where it essentially reproduces EOM-CCSDT's 0.7 eV error)
show that further development will be needed to tackle doubly excited
states more generally.


Active space methods such as complete active space self consistent field
theory (CASSCF) \cite{CASSCF_1982,CASSCF_1985-1,CASSCF_1985-2,CASSCF_1987}
and its second order perturbation theory (CASPT2)
\cite{anderssonSecondorderPerturbationTheory1990,10.1063/1.462209,doi:https://doi.org/10.1002/9781119417774.ch10}
often offer accurate alternatives for modeling doubly excited states,
\cite{sarkar2022assessing,kossoskiReferenceEnergiesDouble2024}
but they are not without their own challenges.
First and foremost, both methods' accuracy can be sensitive to how
large of an active space is chosen.
In principle, the larger the better, but in practice large active
spaces have multiple downsides: most active space solvers display
exponential cost growth with the number of active orbitals,
and the difficulty of maintaining the same qualitative active
space (and thus smooth curves) when scanning a potential energy
surface gets harder as more orbitals are included.
Both effects make minimal active space choices attractive,
such as taking only the hole and particle orbital in the case
of a single-CSF double excitation.
However, as we shall see in our results, this choice can make for
meaningful differences in accuracy, a reality that stems at least
in part from the fact that CASPT2's correlation treatment for electron
pairs unrelated to the active space is essentially that of MP2.


In the ASCC approach, the idea is to use exponentiated de-excitation
operators to initiate a zeroth order coupled cluster
wave function that contains only the excited state's leading CSFs
--- similar to a minimal active space wave function ---
and then to refine the correlation treatment by solving for
the first order coupled cluster amplitudes at a similar $O(N^6)$ cost as seen in typical
ground state CC methods.
This approach has already proven successful in singly excited
states, and in particular in charge transfer states.
\cite{tuckmanAufbauSuppressedCoupledCluster2024,tuckmanImprovingAufbauSuppressedCoupled2025}
Similar to multi-level CC, \cite{myhre2014multi,folkestad2020equation}
ASCC can be adapted so that a high level treatment of an excitation's
most important orbitals is nested within a low level treatment of
the rest, allowing for its scaling to drop to $O(N^5)$ and for
it to reach system sizes much larger than can
be treated with traditional singles and doubles theory.
\cite{tuckmanFastAccurateCharge2025}
Since ASCC has proven to be highly insensitive to the choice of
the molecular orbital (MO) basis, 
\cite{quadyAufbauSuppressedCoupledCluster2025}
it becomes tempting to think that, if it could be generalized to
work for doubly excited states, it may be able to tolerate
relatively straightforward starting points, such as
minimal-active-space state-averaged CASSCF.
In this study, we provide the theory for just such a generalization
and test its accuracy in various single-CSF doubly excited states.
We also perform tests in one preliminary category of multi-CSF
doubly excited state in a proof of principle demonstration that
the approach is not limited to the single-CSF regime.

\section{Theory}
\label{sec:theory}

\subsection{Aufbau Suppression}
\label{sec:theory-as}

In the standard single-reference coupled cluster (CC) theory, the formalism is built upon an exponential ansatz that incorporates the electron correlation effects while maintaining size extensivity. The CC wave function involving a closed shell reference Aufbau determinant $\ket{\Phi_{0}}$ and a cluster operator $\hat{T}$ is represented as
\cite{doi:https://doi.org/10.1002/9780470125915.ch2,doi:https://doi.org/10.1002/9781119019572.ch13}
\begin{align}
    \ket{\Psi_{\textrm{CC}}} = e^{\hat{T}} \ket{\Phi_{0}} = \left(  \mathds{1} + \hat{T} + \frac{1}{2!} \hat{T}^{2} + \ldots
    \right) \ket{\Phi_{0}}
\end{align}
in which $\hat{T}$ is defined as a sum of excitation operators
\begin{align}
    \hat{T}&= \hat{T}_{1}+ \hat{T}_{2} + \ldots + \hat{T}_{n} 
\end{align}
and the $n$-body excitation operator $\hat{T}_{n}$ is given by 
\begin{align}
    \hat{T}_{n}&=   \frac{1}{ \left(  n! \right)^{2}}     \sum_{ij \ldots}^{n} \sum_{ab \ldots}^{n} t_{ij \ldots}^{ab \ldots} \hat{a}^{\dagger}_{a} \hat{a}^{\dagger}_{b} \ldots \hat{a}_{j} \hat{a}_{i}.
\end{align}
Here, the coefficients $t_{ij\ldots}^{ab\ldots}$ are referred to as the cluster amplitudes for the corresponding operators, with $i,j,\ldots$ and $a,b,\ldots$ representing the indices for the occupied and virtual orbitals, respectively. 

To suppress the coefficient on the Aufbau determinant and produce the desired excited state, ASCC introduces a second exponential with a de-excitation operator $\hat{S}^{\dagger}$.
For 1-CSF singly excited states, the resulting ansatz is
\begin{align}
    \ket{\Psi_{\textrm{ASCC}}^{{}^\textrm{1-CSF}}} 
     = e^{-\eta\hat{S}^{\dagger}} e^{\hat{T}} \ket{\Phi_{0}}
    \label{eqn:1CSF-ASCC-wfn}
\end{align}
in which effective choices for $\eta$ and $\hat{S}$ are as follows.
\cite{tuckmanImprovingAufbauSuppressedCoupled2025}
\begin{align}
    \eta &= 1 \\
    \hat{S} & = \frac{1}{\sqrt{2}} \left( \hat{a}^{\dagger}_{p} \hat{a}_{h}
              + \hat{a}^{\dagger}_{\bar{p}}  \hat{a}_{\bar{h}} \right) \\
            \implies \hat{S}^{\dagger} \hat{S} &= \mathds{1}.
\end{align}

Here, $h$ labels a spin-up hole orbital, defined as one that
is occupied in the Aufbau determinant but unoccupied in at least
one of the primary (i.e., large-coefficient)
configurations of the reference state we are attempting to
add a weak correlation treatment to.
Likewise, $p$ labels a spin-up particle orbital, defined as one
that is unoccupied in the Aufbau determinant but occupied in at
least one of the reference's primary configurations.
Orbital labels with an overbar (e.g., $\bar{p}$)
indicate spin-down orbitals.

\subsection{Doubly Excited States}
\label{sec:theory-doubly-excited-states}

Now consider an excited state dominated by a double excitation in which
both electrons transfer between the same hole and particle orbital.
\begin{align}
    \ket{\Psi_{0}}= \ket{\Phi_{h \bar{h}}^{p \bar{p}}}
\end{align}
Our general approach is to produce $\ket{\Psi_{0}}$
(which we will call our reference)
from the Aufbau determinant $\ket{\Phi_{0}}$
(which we refer to as the formal reference)
via careful choices for our de-excitation operator
and the zeroth order parts of $\hat{T}$.
For this 1-CSF doubly excited state, it is enough to propose
$\hat{T}^{(0)} = c_{1} \hat{S} + c_{2} \hat{S}^{2}$,
plug into Eq.\ (\ref{eqn:1CSF-ASCC-wfn}), and solve for the
$c_1$, $c_2$, and $\eta$ values that zero out the Aufbau and
singly excited terms and place a one in front of the doubly
excited term in the wave function expansion.
Doing so results in $c_{1}=\eta=\sqrt{2}$ and $c_2=0$,
which we can verify by plugging back into the ansatz.
\begin{align}
& e^{-\eta\hat{S}^{\dagger}} e^{\hat{T}^{(0)}} \ket{\Phi_{0}} \notag\\
&\qquad = e^{-\sqrt{2}\hat{S}^{\dagger}} e^{\sqrt{2}\hat{S}} \ket{\Phi_{0}} \notag\\
&\qquad =
\left( \mathds{1} - \sqrt{2}\hat{S}^{\dagger} + \hat{S}^{\dagger2} \right)
\left( \mathds{1} + \sqrt{2}\hat{S}  + \hat{S}^{2} \right) \ket{\Phi_{0}}
\notag\\
&\qquad = \left(
\mathds{1} + \sqrt{2}\hat{S}  + \hat{S}^{2}
 - 2 \hat{S}^{\dagger} \hat{S} - \sqrt{2} \hat{S}^{\dagger} \hat{S}^{2}
 + \hat{S}^{\dagger 2} \hat{S}^{2} \right)\ket{\Phi_{0}} \notag\\
&\qquad = \hat{S}^{2} \ket{\Phi_{0}}
        = \ket{\Phi_{h \bar{h}}^{p \bar{p}}}
        = \ket{\Psi_{0}}
\end{align}
Following our approach in singly excited states,
\cite{tuckmanImprovingAufbauSuppressedCoupled2025}
now that we have identified a zeroth order initialization that reproduces
the desired reference, we freeze the $\eta$ value and
set about adding the weak correlation details
by optimizing the zeroth and first order parts of $\hat{T}$.

To this end, we make the usual CC assumption that our ansatz
is an eigenstate of the Hamiltonian and then convert the eigenvalue
equation into a similarity-transformed form.
\begin{align}
    \hat{H} \ket{\Psi_{\textrm{ASCC}}} &= E \ket{\Psi_{\textrm{ASCC}}} \\
    e^{-\hat{T}} {\bar{H}} e^{\hat{T}} \ket{\Phi_{0}} &= E \ket{\Phi_{0}} \label{eq:aa} \\
    \bar{H} &= e^{\eta\hat{S}^{\dagger}} \hat{H} e^{-\eta\hat{S}^{\dagger}}
\end{align}
The energy and amplitude equations needed for our perturbative
analysis and to solve for $\hat{T}$ are then obtained by
projecting our similarity-transformed eigenvalue problem against
the Aufbau determinant $\bra{\Phi_{0}}$ and various excitations $\bra{\mu}$. 
\begin{align}
    \bra{\Phi_{0}} e^{-\hat{T}} \bar{{H}} e^{\hat{T}} \ket{\Phi_{0}} &= E \\
    \bra{\mu} e^{-\hat{T}} \bar{H} e^{\hat{T}} \ket{\Phi_{0}} &= 0 \label{eq:mm}
\end{align}


As a side note, we should emphasize the similarities and
differences that these equations' double similarity transforms have
with those of the
extended coupled-cluster method (ECCM).
\cite{PhysRevA.36.2519,Fan10082005,VANVOORHIS2000585}
Both approaches have one similarity transform with an excitation operator
and one with a de-excitation operator.
However, ASCC employs a much more limited de-excitation operator, allowing
it to maintain CCSD's $O(N^6)$ cost scaling, as compared to the 
$O(N^{10})$ of untruncated ECCM (although see very recent work
from Loos and coworkers \cite{loos_arXiv_GW_ECC_2026}).
The role of the de-excitation transform is also different:
in ASCC it is used to set up an excited state, whereas in ECCM
its role is to make ground state CC more closely resemble a variational form.


Before we get into the details of choosing the first order part of $\hat{T}$,
it is worth considering whether a similar zeroth order setup can be achieved for
multi-configurational double excitations.
While we will not seek to find a general solution to multi-configurational
excitations in this initial application of ASCC to doubly excited states,
let us at least crack open the door via a demonstration on 2-CSF
double excitations that involve two hole orbitals and one particle orbital,
such as can be found in glyoxal and similar molecules.
The reference for such states can be written as
\begin{align}\label{eq:dd}
    \ket{\Psi_{0}} &= a \ket{\Psi_{h_{1} \bar{h}_{1}} ^{p_{1} \bar{p}_{1}}} + b \ket{\Psi_{h_{2} \bar{h}_{2}} ^{p_{1} \bar{p}_{1}}} 
\end{align}
where normalization requires that $a^{2}+b^{2}=1$.
To handle this case, we choose a slightly more involved
exponentiated de-excitation operator.
\begin{align}
\ket{\Psi_{\textrm{ASCC}}^{{}^\textrm{2-CSF}}} 
    &= e^{-\gamma\hat{S}^{\dagger}_{11} 
          -\omega\hat{S}^{\dagger}_{12}} \hspace{0.7mm}
       e^{\hat{T}} \hspace{0.7mm} \ket{\Phi_{0}}
    \label{eqn:2CSF-ASCC-wfn} \\
    \hat{S}_{11} &= \frac{1}{\sqrt{2}} \left(
                      \hat{a}^{\dagger}_{p_{1}} \hat{a}_{h_{1}} +
                      \hat{a}^{\dagger}_{\bar{p}_{1}} \hat{a}_{\bar{h}_{1}}
                    \right) \\
    \hat{S}_{12} &= \frac{1}{\sqrt{2}} \left(
                      \hat{a}^{\dagger}_{p_{1}} \hat{a}_{h_{2}} +
                      \hat{a}^{\dagger}_{\bar{p}_{1}} \hat{a}_{\bar{h}_{2}}
                    \right)
\end{align}
As before, the zeroth order part of $\hat{T}$ involves
only the primary orbitals, which in this case are $h_1$, $h_2$, and $p_1$
We start with the following general form involving the single and double
excitations within these orbitals.
\begin{align}\label{eq:ee}
    \hat{T}^{(0)}=  c_{1} \hat{S}_{11} + c_{2} \hat{S}_{11}^{2}
                  + c_{3} \hat{S}_{12} + c_{4} \hat{S}_{12}^{2} + c_{5} \hat{S}_{11} \hat{S}_{12}.
\end{align}
The goal is to initialize $\hat{T}^{(0)}$ so that, at zeroth order,
our ansatz matches the leading CSFs from the CASSCF reference, which in
this two-CSF case means we want to match Eq.\ (\ref{eq:dd}).

We therefore plug $\hat{T}^{(0)}$ in for $\hat{T}$ and solve for the
$\gamma$, $\omega$, and $c_1$ through $c_5$ values that
turn $\ket{\Psi_{\textrm{ASCC}}^{{}^\textrm{2-CSF}}}$ into $\ket{\Psi_0}$.
As an aside, we note that the algebra is more straightforward if we
move the de-excitation exponentials to the side of the equation
with $\ket{\Psi_0}$, so that we are solving
\begin{align}
    e^{\hat{T}^{(0)}} \ket{\Phi_0} =
    e^{ \gamma\hat{S}^{\dagger}_{11} + \omega\hat{S}^{\dagger}_{12} } \ket{\Psi_0},
\end{align}
which requires that the coefficients on the Aufbau, the two singly excited
configurations, and the three doubly excited configurations match in the
right hand side's and left hand side's wave function expansions.
Note that, in this example, neither side has any triple or higher excitations,
as we cannot put three electrons in the particle orbital.
We also enforce $\gamma^2 + \omega^2 = 1$ to ensure the magnitude of
the overall de-excitation operator remains normalized, as it was in
the 1-CSF case. Solving this system of seven equations produces
$\gamma$, $\omega$, and $c_1$ through $c_5$ values that, for the
specific $a$ and $b$ coefficients in question,
ensure that the zeroth order part of
$\ket{\Psi_{\textrm{ASCC}}^{{}^\textrm{2-CSF}}}$
gets initialized to $\ket{\Psi_0}$.
At this point, we freeze the $\gamma$ and $\omega$ values
and move forward to fine-tuning the zeroth and first order
parts of $\hat{T}$
via Eq.\ (\ref{eq:mm}).

\subsection{Truncating the Cluster Operator}
Before delineating the details of how a perturbative treatment
guides our truncation of the cluster operator in ASCC, let us briefly review
how this type of treatment works in ground state CC.
First, we partition the Hamiltonian as $\hat{H}=\hat{F}+\hat{V}$,
where we define the
Fock operator $\hat{F}$ as the zeroth order part and
the two-body operator $\hat{V}$ as the first order correction.
These definitions allow every term in the
amplitude update equations to be assigned an order relative to the
this Hamiltonian correction.
For example, one finds that the largest contribution to $\hat{T}_2$
looks like $\hat{V} / \Delta$, where $\Delta$ is a difference
between diagonal entries of the zeroth order Fock matrix.
Thus, one concludes that $\hat{T}_{2}$ is first order.
If, as is the case for us here, Brillouin's condition is not satisfied,
$\hat{T}_{1}$ is also found to be first order.
In the ground state theory, all other parts of $\hat{T}$ are found to
be second order or smaller, and so one way to motivate the CCSD
ansatz is that it truncates $\hat{T}$ to its first order components.
For ASCC, we pursue the exact same strategy, but with a slightly more
involved choice for the zeroth order Hamiltonian that reflects the
special nature of the hole and particle orbitals.



\subsubsection{Perturbative Setup}

Truncating $\hat{T}$ to its zeroth and first order parts can be achieved
through the same form of perturbative analysis used for singly excited ASCC,
\cite{tuckmanImprovingAufbauSuppressedCoupled2025}
which in turn is similar to how perturbative arguments support
common truncations of $\hat{T}$ in ground state CC theory.
\cite{doi:https://doi.org/10.1002/9780470125915.ch2,doi:https://doi.org/10.1002/9781119019572.ch14}
In the ground state theory, there are no zeroth order amplitudes
within $\hat{T}$, and so the situation in ASCC is a bit more involved.
To start, we define the zeroth and first order parts of the Hamiltonian as follows.
\begin{align}
    \label{eqn:0thham}
    \hat{H}^{(0)} &= \hat{F}^{(oo)} +
                     \hat{F}^{(vv)} +
                     \hat{F}^{(hh)} +
                     \hat{F}^{(pp)} \\
    \label{eqn:1stham}
    \hat{H}^{(1)} &= \hat{H} - \hat{H}^{(0)} 
\end{align}
Here, $\hat{F}$ is the one-body Fock operator in the MO basis,
and $\hat{F}^{(oo)}$ and $\hat{F}^{(vv)}$ are its components in the subspaces spanned by non-hole occupied orbitals and non-particle virtual
orbitals, respectively.
Similarly,
$\hat{F}^{(hh)}$ and $\hat{F}^{(pp)}$ are the parts of the Fock operator
that map within the span of the hole orbitals
and within the span of the particle orbitals, respectively.
Note that we do not restrict $\hat{F}^{(oo)}$, $\hat{F}^{(vv)}$, 
$\hat{F}^{(hh)}$, or $\hat{F}^{(pp)}$ to be diagonal in this formal
perturbative analysis.

In addition to breaking up the zeroth order Hamiltonian, we also find it
helpful to separate the cluster operator $\hat{T}$ into three parts
\begin{align}
    \label{eqn:t-breakup}
    \hat{T} &= \hat{T}^N + \hat{T}^M + \hat{T}^P
\end{align}
and to reorganize our ansatz as
\begin{align}
    \label{eqn:breakout}
    \ket{\Psi_{\textrm{ASCC}}} &= e^{\hat{T}^{N}} e^{-\hat{Z}^{\dagger}} e^{\hat{T}^{M}} e^{\hat{T}^{P}} \ket{\Phi_{0}},
\end{align}
where $\hat{Z}^\dagger$ contains the de-excitation operators:
$\eta\hat{S}^\dagger$ in 1-CSF ASCC and
$\gamma\hat{S}^\dagger_{11}+\omega\hat{S}^\dagger_{12}$
in our 2-CSF example. It is worth noting that de-excitation operators within $\hat{Z}^{\dagger}$ are zeroth order, and therefore $\hat{Z}^{\dagger}$ itself is zeroth order. This breakup of the cluster operator allows us to appreciate
the mathematical and physical roles played by its different parts.
The ``non-primary''
$\hat{T}^{N}$ operator contains all parts of the cluster operator
that have no particle or hole indices (i.e., they have only ``non-primary,''
or, in other words, ``external'' indices)
which allows it to commute with the de-excitation operator and get
pulled out in front.
This positioning emphasizes the fact that $\hat{T}^N$ will function much
like a cluster operator in many multi-reference CC theories.
For example, in our 2-CSF tests below,
$\exp(\hat{T}^N)$ acts on the multi-configurational reference state
that gets set up by the interaction of
$\hat{T}^{P}$ and $\hat{Z}^\dagger$.
The other two components, $\hat{T}^{M}$ and $\hat{T}^{P}$, do
not commute with $\hat{Z}^\dagger$, because
$\hat{T}^{M}$ is the ``mixed'' part,
defined as containing all components within $\hat{T}$ that
have both non-primary and primary indices, while $\hat{T}^{P}$
is the ``primary'' part,
defined as containing all components that have only primary indices.


\subsubsection{Zeroth Order}

With the ansatz written this way, we can quickly confirm that $\hat{T}^P$
contains $\hat{T}^{(0)}$ and that the largest components of
$\hat{T}^{N}$ and $\hat{T}^{M}$ are first order or smaller.
To see why, start by noting that,
in the zeroth order simplification of Eq.\ (\ref{eq:mm}),
our block-diagonal $\hat{H}^{(0)}$
cannot move electrons between primary
and non-primary orbitals.
\begin{align}
    \label{eqn:0th-proj}
    \bra{\mu} e^{-\hat{T}^{(0)}} e^{\hat{Z}^\dagger}
    \hat{H}^{(0)}
    e^{-\hat{Z}^\dagger}
    e^{\hat{T}^{(0)}} \ket{\Phi_{0}} &= 0
\end{align}
Thus, to zeroth order, we can satisfy
the all-non-primary $\bra{\mu^N}$ and
the mixed $\bra{\mu^M}$ projection equations
by holding the zeroth order parts of
$\hat{T}^{N}$ and $\hat{T}^{M}$ at zero.
Doing so, the all-primary $\bra{\mu^P}$ projection equations
can then be solved to zeroth order by adjusting the amplitudes
within $\hat{T}^P$, which we need to remember are nonzero
to start with due to our wave function initialization.
In practice, we do not actually solve for $\hat{T}^{(0)}$ in isolation.
Instead, we use this analysis to recognize that all of
$\hat{T}^{(0)}$ is contained in $\hat{T}^P$, which helps
justify our practical setup in which our initial guess
initializes $\hat{T}^P$ as discussed in
Sections \ref{sec:theory}\ref{sec:theory-as}
and \ref{sec:theory}\ref{sec:theory-doubly-excited-states} while
simply initializing $\hat{T}^M$ and $\hat{T}^N$ to zero.


\subsubsection{Mixed Amplitudes with 4+ Non-Primaries}

In our quest to include only zeroth and first order amplitudes
in our truncated ansatz,
the first batch of amplitudes that we will rule out are those
within the mixed $\hat{T}^M$ operator that have four or more non-primary indices.
To do so, we consider the first order part of the mixed projection
equations,
\begin{align}
    \label{eqn:1st-order-M-proj}
    \bra{\mu^M} & \hspace{0.7mm} \left[ \mathring{H}^{(0)}, \hat{T}^{(1)} \right]
      + \mathring{H}^{(1)} \hspace{0.7mm} \ket{\Phi_0} = 0,
\end{align}
where we employ the shorthand
\begin{align}
   \mathring{H}^{(0)} &= e^{-\hat{T}^{(0)}} e^{\hat{Z}^\dagger}
                  \hat{H}^{(0)} \hspace{0.7mm}
                  e^{-\hat{Z^\dagger}} e^{\hat{T}^{(0)}}  \\
        \mathring{H}^{(1)} &= e^{-\hat{T}^{(0)}} e^{\hat{Z}^\dagger}
                  \hat{H}^{(1)} \hspace{0.7mm}
                  e^{-\hat{Z^\dagger}} e^{\hat{T}^{(0)}}    
\end{align}
to denote similarity transforms of $\hat{H}^{(0)}$ and $\hat{H}^{(1)}$ with respect to the
de-excitation operator and zeroth order part of the cluster operator.
We start by noting that the only way for the $\mathring{H}^{(1)}$
term to make a nonzero contribution when $\bra{\mu^M}$ contains
four or more non-primary indices is via the
all-non-primary  part of the two-electron operator in $\hat{H}$ itself, because all the indices on
$\hat{T}^{(0)}$ and $\hat{S}^{\dagger}$ are primary indices. However, this part of $\hat{H}$ commutes with $\hat{Z}^{\dagger}$
and $\hat{T}^{(0)}$, and so the corresponding component of
$\mathring{H}^{(1)}$ will only contain external excitations and will thus
be unable to connect a mixed $\bra{\mu^M}$ (which also contains one or more
primary excitations) with $\ket{\Phi_0}$.
Therefore, when a projection is mixed and involves four or more non-primary indices, only the $[ \mathring{H}^{(0)}, \hat{T}^{(1)} ]$ commutator
contributes. 
With $\mathring{H}^{(0)}$ unable to move electrons between primary
and non-primary orbitals, any contraction it makes with
an operator inside $\hat{T}^{(1)}$ will produce a new operator
with the same number of primary and non-primary indices
as the original operator.
Thus, the only terms in the commutator that can connect
$\bra{\mu^M}$ with $\ket{\Phi_0}$ when $\mu^M$
has four or more non-primary indices
are those whose amplitudes also have four or more non-primary indices.
Thus, setting those amplitudes to zero solves these projections
to first order, and we conclude that such amplitudes are second
order or smaller.

\subsubsection{All-Non-Primary Doubles}

Turning our attention to the first order part of the
all-non-primary projections,
\begin{align}
    \label{eqn:1st-order-N-proj}
    \bra{\mu^N} & \hspace{0.7mm} \left[ \mathring{H}^{(0)}, \hat{T}^{(1)} \right]
      + \mathring{H}^{(1)} \hspace{0.7mm} \ket{\Phi_0} = 0,
\end{align}
we can conclude that the doubles amplitudes
inside $\hat{T}^N$ will be first order, and, further, that all
higher excitations within $\hat{T}^N$ will be second order or smaller.
Looking at the doubles subset of the $\bra{\mu^N}$ projections,
we see that each such projection gets a first order contribution
from $\mathring{H}^{(1)}$.
For a mixed amplitude from $\hat{T}^M$ to contribute to this projection
via the commutator term, it would have to have four non-primary indices.
As we discussed in the previous section, these mixed amplitudes
are second order or smaller.
Thus, the only amplitudes left that can make a first order contribution via
contractions with $\mathring{H}^{(0)}$ are the doubles within $\hat{T}^N$,
leading us to conclude that these doubles are first order.
Similarly, the triples and higher
within $\hat{T}^N$ are second order or smaller,
because (i) they are the only amplitudes that contribute to their own projection
equations at first order, and (ii) $\mathring{H}^{(1)}$ makes no contribution
to those projections as it cannot produce more than two non-primary excitations.

\subsubsection{Other Amplitudes}

Finally, we consider the non-primary singles amplitudes and the
various different categories of mixed amplitudes that have one,
two, or three non-primary indices.
Due to the ability of $\mathring{H}^{(0)}$ to perform a de-excitation
in the primary space, many of these amplitudes show up at first order
in each other's amplitude equations, making things more complicated than
in standard ground state theory.
For example, consider the set of amplitudes $\{t| \substack{a\\i}\}$,
which contains all amplitudes that have the non-primary indices
$i$ and $a$, as well as zero or more primary indices.
In 1-CSF ASCC as well as our 2-CSF example, this set will contain
the non-primary singles amplitude $t_i^a$ and various mixed doubles
and triples amplitudes (e.g., $t_{ih}^{ap}$ and $t_{ih\bar{h}}^{ap\bar{p}}$).
Some of the doubles amplitudes in this set contribute
to the $\bra{\substack{a\\i}}$ projection via the
$[\mathring{H}^{(0)},\hat{T}^{(1)}]$ commutator due to terms
involving the de-excitation operator.
More familiarly, at least from the perspective of ground state theory,
the $t_i^a$ amplitude also contributes to this projection at first order
via its contraction with $\hat{H}^{(0)}$.
Thus, the $\bra{\substack{a\\i}}$ projection gets first order contributions
from multiple different types of amplitudes within $\{t| \substack{a\\i}\}$.
Similarly, the doubles in this set contribute to their own doubles
projections (e.g., $\bra{\substack{ap\\ih}}$), but these 
projections also get contributions from the triples in the set,
again via de-excitation from the $[\mathring{H}^{(0)},\hat{T}^{(1)}]$
commutator.
The set's triples amplitudes have corresponding projections
(e.g., $\bra{\substack{ap\bar{p}\\ih\bar{h}}}$) that, somewhat
counterintuitively if you are coming from the perspective of ground state CC,
get non-zero first order contributions from the $\mathring{H}^{(1)}$
term due to the commutator between the two-electron part of $\hat{H}$
and the all-primary doubles within $\hat{T}^{(0)}$
(which are not necessarily zero even if we initialize them that way,
as their own amplitude equations have a nontrivial zeroth order part
as discussed above).
Thus, the projections corresponding to the amplitudes in our
$\{t| \substack{a\\i}\}$ set all get first order contributions from
$\mathring{H}^{(1)}$ and from one or more
of the set's amplitudes via the $[\mathring{H}^{(0)},\hat{T}^{(1)}]$
commutator.
Note, however, that amplitudes that are not in the set
do not contribute to these equations at first order,
and so the equations form a closed system
containing one equation per amplitude.

The conclusion is that all the amplitudes in $\{t| \substack{a\\i}\}$
are first order.
Similar reasoning reaches the same conclusion for the amplitude
sets $\{t|\substack{ab\\i}\}$, $\{t| \substack{a\\ij}\}$,
$\{t| \substack{\vphantom{a}\\i}\}$, $\{t| \substack{a\\ \vphantom{i} }\}$,
$\{t| \substack{\vphantom{a}\\ij}\}$, and $\{t| \substack{ab\\ \vphantom{i} }\}$.
However, as discussed above, mixed amplitudes in $\{t|\substack{ab\\ij}\}$
and other sets with more than three non-primary indices are second order
or smaller.
Likewise, the amplitudes in $\{t|\substack{abc\\ \vphantom{i}}\}$
and $\{t|\substack{\vphantom{a}\\ijk}\}$ are also second order or smaller,
as the two-electron part of $\hat{H}$ inside $\mathring{H}^{(1)}$
cannot create three electrons in the virtuals or three holes in the occupieds.
Bringing it all together, we find that the parts of $\hat{T}$ that
are first order or larger are:
all the singles, all the doubles, and all the triples that
(i) have a primary hole index,
(ii) have a primary particle index,
and (iii) have three or more primary indices overall.
We include these amplitudes in our cluster operator and
neglect the others (which are all second order or smaller),
leaving us with an ASCCSD theory containing an overall number
of amplitudes that is asymptotically identical to ground state CCSD
(since we have only $O(N^3)$ triples amplitudes).

\subsubsection{Multi-Configurational Considerations}

In both our 1-CSF approach and our 2-CSF example,
there is only one primary particle orbital, which implies that
$\hat{T}^P$ and thus $\hat{T}^{(0)}$ will only contain
single and double excitations,
because it is not possible to excited more than two electrons into one orbital.
Furthermore, it implies that the first order subsets of the
amplitude sets (e.g., $\{t| \substack{a\\i}\}$) discussed in the
previous section top out at triple excitations.
An interesting question, although one beyond the scope of
this study, is whether all excitation levels in the
primary space would need to be included in
situations where there were both multiple primary hole
and multiple primary particle orbitals.
A preliminary perturbative analysis suggests that the answer is yes,
which would limit the approach taken here to relatively
small primary spaces.
We estimate that two particles and two holes would be doable
(indeed it has been done for singly excited
states \cite{tuckmanImprovingAufbauSuppressedCoupled2025}),
because there is only one quadruple inside that modest primary space.
The three-particle/three-hole case would be more challenging,
and beyond that computer memory would likely become prohibitive,
because the $O(N^3)$ storage needs of the amplitude sets
$\{t| \substack{ab\\i}\}$ and $\{t| \substack{a\\ij}\}$ would now be
multiplied by a factor that is exponential in the number of primary orbitals.
So, we estimate that the present approach will be able
to extend to somewhat more complicated multi-configurational
doubly excited states than what we test in this study,
but that further innovations will be necessary
to tackle cases with large numbers of primary orbitals.

\subsubsection{Comparison to Singly Excited ASCC}

Interestingly, for 1-CSF cases, the set of amplitudes that
this perturbative analysis retains is the same for
doubly excited states as it is for singly excited states.
From a mathematical perspective, 1-CSF singly excited ASCC
and 1-CSF doubly excited ASCC end up being different solutions
to the same set of nonlinear equations.
The key differences come from differing numerical factors
in front of the de-excitation operator and in the initial guess
values used for the all-primary amplitudes.
Indeed, since the perturbative analysis carried out here mirrors
that of 1-CSF singly excited states,
\cite{tuckmanImprovingAufbauSuppressedCoupled2025}
perturbative improvements and cost saving measures
(such as nesting\cite{tuckmanFastAccurateCharge2025}) that
work for one theory should work for both. 
Another consequence of these theories essentially sharing their
working equations is that they also share computational costs.
In particular, the asymptotic cost of 1-CSF doubly excited ASCC,
like that of its singly excited cousin,
\cite{tuckmanImprovingAufbauSuppressedCoupled2025}
is the same as CCSD.
The glyoxal-inspired 2-CSF incarnation of doubly excited ASCC
does have more expensive sub-dominant terms due to the
larger range of its hole indices,
but its $O(N^6)$-cost terms are identical to those of the 1-CSF
ASCC theories and to those of CCSD.
Given the challenging history of double excitations in
coupled cluster theory, especially those containing more than one
important CSF, it would be quite noteworthy if a theory with CCSD's
asymptotic cost could treat them accurately.
To find out, let's take a look at some results.

\section{Results}
\subsection{Computational Details}
%

\begin{figure*}[t]
    \centering
    \includegraphics[width=1.0\linewidth]{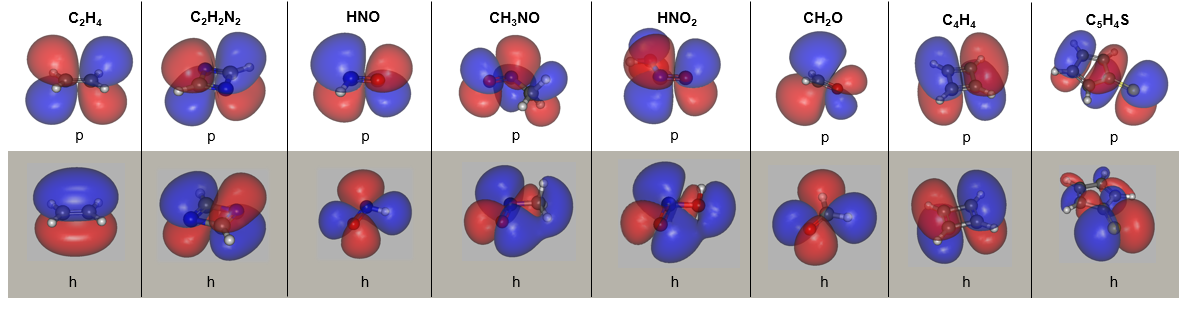}
    \caption{The primary hole (h) and particle (p) orbitals of the CASSCF reference for the 1-CSF doubly excited states.}
    \label{fig:1csf-orbs}
\end{figure*}
\begin{figure*}[t]
    \includegraphics[width=1.0\linewidth]{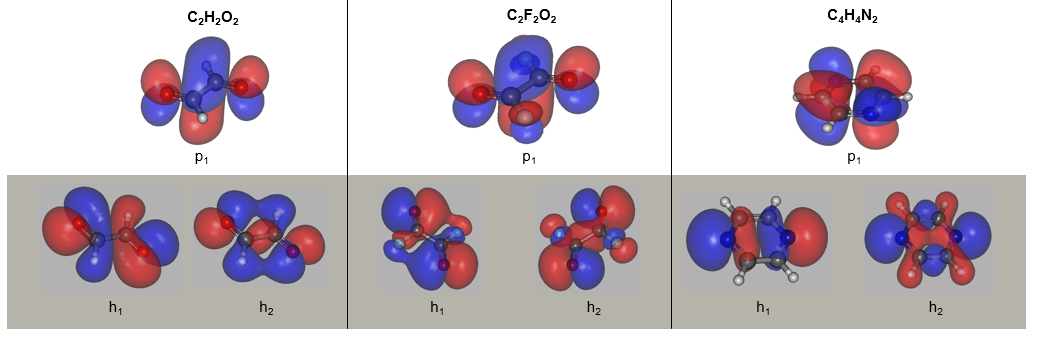}
    \caption{The primary hole (h) and particle (p) orbitals of the (3o,4e) CASSCF reference for the 2-CSF examples.}
    \label{fig:2csf-orbs}
\end{figure*}

EOM-CCSD calculations were performed with Q-Chem.\cite{10.1063/5.0055522} CASPT2 \cite{Andersson1990CASPT2} calculations were carried out using MOLPRO\cite{https://doi.org/10.1002/wcms.82,10.1063/5.0005081}. EOM-CC3\cite{RevModPhys.79.291,10.1063/1.473322}, EOM-CCSDT\cite{10.1063/1.467620,10.1063/1.1416173}, and reference excitation energies come from the QUEST \#2 benchmark dataset. \cite{kossoskiReferenceEnergiesDouble2024} These excited-state calculations were performed using the frozen-core approach, whereas ASCC and EOM-CCSD calculations did not use the frozen-core approximation. Given that the frozen core approximation only slightly alters the excitation energies ($\sim$0.02 eV), \cite{kossoskiReferenceEnergiesDouble2024,doi:10.1021/acs.jctc.8b01205} it is reasonable to compare between methods that freeze and do not freeze the core.
Molecular geometries are also taken from the QUEST \#2 benchmark set,
\cite{kossoskiReferenceEnergiesDouble2024}
where they were optimized at CC3/aug-cc-pVTZ level of theory.
Multiple basis sets are tested for ASCC, including 6-31+G$^{*}$ (with spherical d functions),
aug-cc-pVDZ, and aug-cc-pVTZ.

For ASCC, the doubly excited reference states were determined via
CASSCF calculations in PySCF.
\cite{https://doi.org/10.1002/wcms.1340,10.1063/5.0006074}
Specifically, the ASCC starting point employs the
CASSCF canonical molecular orbital basis and a truncated CASSCF wave function:
only CSFs with coefficients above 0.25 are retained.
Equal weight state-averaged CASSCF calculations with an active space of 2
electrons in 2 orbitals, CASSCF(2o,2e), were used for single-CSF states with the ground and doubly excited states given equal weights.  
For our two-CSF examples, the same approach was taken but the active
space was increased to (3o,4e) to cover both the particle orbital
and the two different hole orbitals.
While in most cases this setup led to convergent ASCC calculations,
two molecules proved to be exceptions.
In formaldehyde (in all three basis sets) and in diazete
(in the 6-31+G$^{*}$ basis), ASCC did not converge when started from
state averaged CASSCF.
In these cases, we found that ASCC does converge when
started from state-specific CASSCF wave functions, which were evaluated
by setting the state averaging weights to zero on the lower states
(no root flipping issues were observed).
Thus, the ASCC results presented below for these four cases are based
on state-specific starting points, while all others are based on
equal-weight state average starting points.
The convergence thresholds for the ASCC energy and amplitude updates were
set at $10^{-7}$ a.u.\ and $10^{-5}$, respectively.

\subsection{Single-CSF Tests}

We begin by looking at a collection of single-CSF double excitations (with the orbitals involved in excitation shown in Figure \ref{fig:1csf-orbs}),
all of which have a CC3/AVTZ $\%T_{1}$ metric below 50\%.
As seen in Table \ref{tab:ASCCtable}, and as noted before, \cite{10.1063/5.0091715}
EOM-CCSD performs quite poorly for these states,
with typical excitation energy errors in the 2 eV to 4 eV range.
Higher levels of EOM theory reduce this error systematically,
but it is not until one gets to EOM-CC4 (as reported in QUEST \# 2 benchmark\cite{kossoskiReferenceEnergiesDouble2024}) that typical errors fall below 0.2 eV.
These errors stand in stark contrast to the high accuracy that EOM methods
achieve for singly excited states, and are especially noteworthy given the
high cost of this systematic improvement:
EOM-CCSD, EOM-CC3, EOM-CCSDT, and EOM-CC4 have costs that
grow as the 6th, 7th, 8th, and 9th power of the system size, respectively.
In comparison, (2o,2e)-CASSCF-based ASCC achieves a mean unsigned error
in single-CSF states below 0.2 eV already at the singles and doubles level,
for which the cost grows as the 6th power of the system size.
Whether we categorize the error by mean signed error, mean unsigned error,
or maximum error, this set of single-CSF double excitations shows the same
accuracy ordering: singles-and-doubles ASCC is more accurate than EOM-CCSDT, which in turn out-performs all lower levels of EOM-CC. Interestingly, ASCC delivers comparable accuracy for 1-CSF singly excited and 1-CSF doubly excited states, which is rarely achieved by excited state methods. 
Thus, by explicitly centering its wave function expansion on the
doubly excited reference determinant, ASCC achieves the hoped for
advantage over methods that linearly respond around the ground state. Excited state specific methods, such as $\Delta$CC and the intermediate state approach to EOM-CC (IS-CC), achieve something similar
for single-configurational doubly excited states, as shown in Figures \ref{fig:deltas} and \ref{fig:intermediates}.


Before turning to those tests, we would like to make a small aside
about ASCC's lack of sensitivity to the initial orbitals.
Previous work has already observed that, for singly excited states,
the choice of the reference method that generates the molecular orbital
shapes typically has only a small effect on the ASCC excitation energy.
\cite{quadyAufbauSuppressedCoupledCluster2025}
To perform a limited test of this property for double excitations,
we have re-calculated the ASCC excitation energies using equal weight state-averaged
CASSCF(4o,4e) reference in both diazete and cyclobutadiene
and then compared them to the ASCC results based on CASSCF(2o,2e).
In both moleules, we find that this change to the active space
changes the ASCC excitation energy 
by less than 0.07eV.

\subsection{Two-CSF Tests}

In the glyoxal molecule, there are two different oxygen lone pair
orbitals whose $n^2\rightarrow (\pi^*)^2$ transitions are close
enough in energy to mix strongly into a relatively low-lying
two-CSF doubly excited state.
Specifically, as seen in Figure \ref{fig:2csf-orbs},
each oxygen has a lone pair that, more or less,
occupies the non-bonding 2p orbital perpendicular
to the CO bond, and these ``local'' lone pairs combine
to make one asymmetric and one symmetric molecular orbital,
which we label as $h_1$ and $h_2$, respectively.
The same basic pattern in which two degenerate local lone
pairs mix into symmetry-respecting molecular orbitals
is also seen
in pyrazine and oxalyl fluoride, and in all three molecules
the result is a doubly excited state with large contributions
from both the $h_1^2\rightarrow (\pi^*)^2$ and
$h_2^2\rightarrow (\pi^*)^2$ CSFs.
In contrast to the single-CSF states discussed in the previous
section, these states are difficult to treat accurately with
$\Delta$CC, \cite{doi:10.1021/acs.jctc.4c00034}
and the errors they engender in EOM methods are even more severe.

\begin{figure}
    \centering
    \includegraphics[width=0.86\linewidth]
    {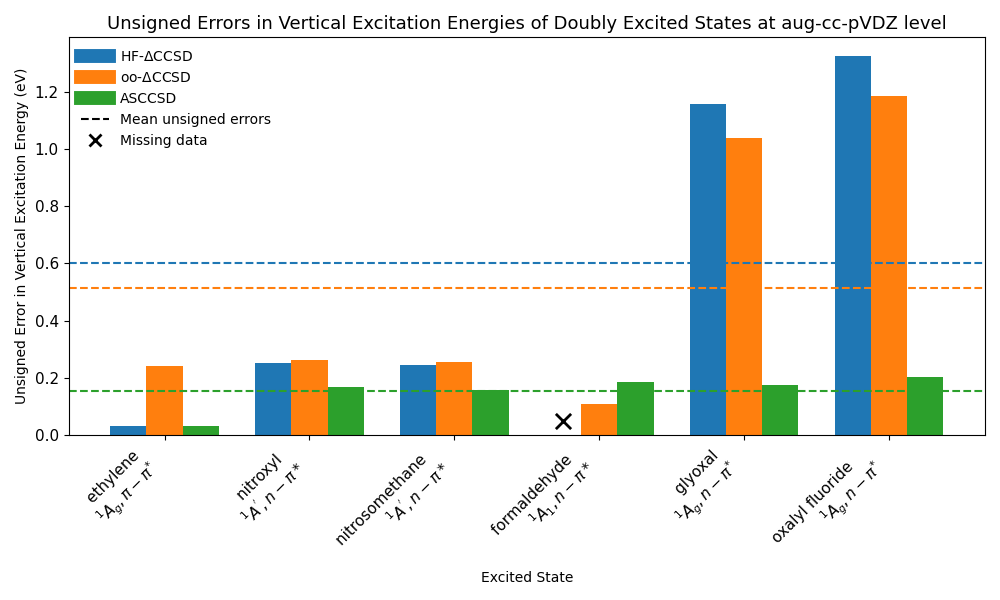}
    \captionsetup{labelfont={bf,color=black},textfont={color=black}}
    \caption{\color{black}Unsigned errors (in eV) in vertical excitation energies of doubly excited states in aug-cc-pVDZ.
    ASCC is compared against recently reported results
    \cite{doi:10.1021/acs.jctc.4c00034}
    for two $\Delta$CC methods: HF-$\Delta$CCSD and oo-$\Delta$CCSD.
    Errors are relative to QUEST basis-specific theoretical best estimates (TBE). The x denotes the case for which a HF-$\Delta$CCSD result was not reported. \cite{doi:10.1021/acs.jctc.4c00034} }
    \label{fig:deltas}
\end{figure}

\begin{figure}
    \centering
    \includegraphics[width=0.90\linewidth]
    {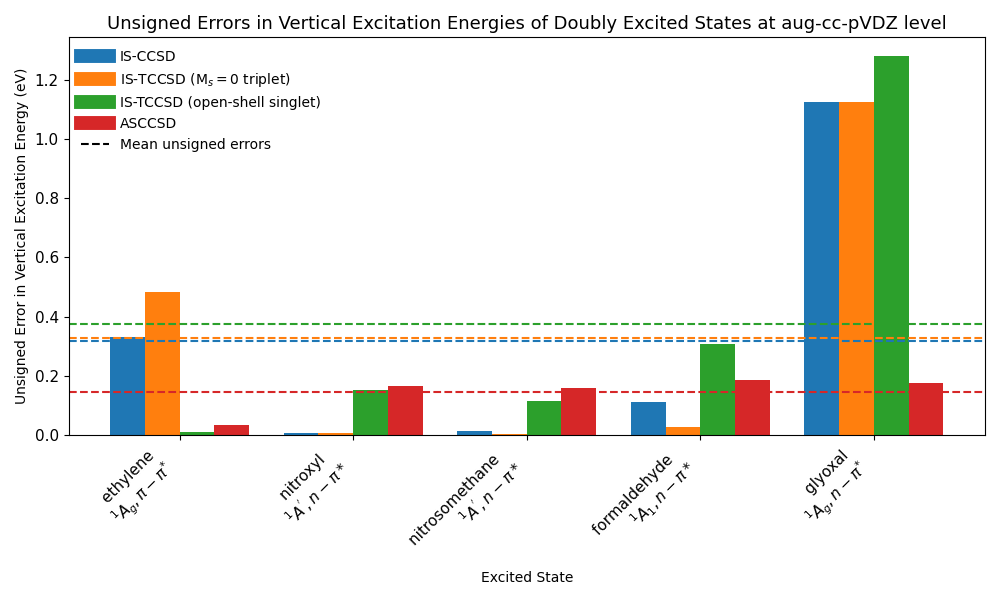}
    \captionsetup{labelfont={bf,color=black},textfont={color=black}}
    \caption{\color{black}Unsigned errors (in eV) in vertical excitation energies of doubly excited states in aug-cc-pVDZ.  ASCC is compared against recently
    reported results \cite{10.1063/5.0091715} for
    IS-CC approaches: IS-CCSD, IS-TCCSD (M$_{\textrm{s}}=0$ triplet) and IS-TCCSD (open-shell singlet). Errors are relative to the QUEST basis-specific theoretical best estimates (TBE). }
    \label{fig:intermediates}
\end{figure}

When started from a CASSCF(3o,4e) reference that has been
truncated down to its two major determinants, ASCC achieves
excitation energy accuracies in these 2-CSF states that
are essentially the same as those for 1-CSF states,
as shown in Table \ref{tab:ASCCtable}
\textcolor{black}{and Figures \ref{fig:deltas} and \ref{fig:intermediates}.}
This result contrasts notably with EOM-CCSD, which shows
even larger errors ($\sim$ 6 eV) for the 2-CSF states
than for the 1-CSF states ($\sim$ 4 eV). 
Even full-triples EOM-CCSDT makes errors above 0.5 eV
for these states, while the errors for ASCC are less than
0.25 eV, with the average unsigned error being just 0.14 eV.
These results appear to confirm that the initialization of the
zeroth order ASCC amplitudes is correctly capturing the
two-CSF nature of the state, and that this fidelity
is maintained after solving the amplitude equations that
fine-tune the weak correlation treatment.

\begin{table*}[ht]
    \centering
     \caption{QUEST TBE for double
     excitation energies in each basis set,
     as well as various methods' errors relative to the basis-specific TBE, all in eV.
     } 
\scriptsize
\begin{tabular}{llllcccccc}
\toprule
\textbf{molecule} & \textbf{state} & \textbf{transition} & \textbf{basis set}  &   \textbf{CSFs} & \textbf{EOM-CCSD}  &  \textbf{EOM-CC3}   & \textbf{EOM-CCSDT}  &  \textbf{ASCCSD} & \textbf{TBE} \\
\midrule
\hline \\
      ethylene  & $^{1}$A$_{g}$ &    $(\pi)^{2} \to (\pi^{*})^{2}$ &  6-31+G*  &  1   &   2.056   &  0.433   &   0.110    &  -0.009 &  13.387\\
         &  &   &  aug-cc-pVDZ &   1  & 2.300  &  0.501   &   0.136  &  0.033  & 13.068 \\
        &   & & aug-cc-pVTZ  &  1   &  3.584   &  0.521   &  0.181  &   0.084     &  12.899 \\
       diazete & $^{1}$A$_{1}$  &   $(\pi)^{2} \to (\pi^{*})^{2}$ & 6-31+G*   &  1   &  4.102     & 0.739  &  0.371   &  0.300    & 6.726 \\
        &  &    & aug-cc-pVDZ    &  1   &   4.340   &  0.820  &  0.445 &  0.285      &  6.635 \\
        &   &    &  aug-cc-pVTZ  &  1  &  4.947   &   0.904   &  0.587   &  0.305     & 6.605  \\
       nitroxyl &   $^{1}$A$'$    &  $(n)^{2} \to (\pi^{*})^{2} $  & 6-31+G*    &    1 & 3.653    &   0.767   &  0.308   &  0.148       &  4.511   \\
         &   &   &  aug-cc-pVDZ     &  1 &   3.848   &  0.850   &   0.359  &  0.167       &  4.397   \\
         &  &   &   aug-cc-pVTZ    &   1  &   4.530    &  0.924    &   0.452 &  0.181      &  4.333  \\
         nitrosomethane &  $^{1}$A$'$   & $(n)^{2} \to (\pi^{*})^{2} $ & 6-31+G*   &        1  &   4.001   &  0.868  &  0.399   &  0.147     &  4.861 \\
          &    &     & aug-cc-pVDZ     & 1  &  4.265    &  0.933  &   0.442    &  0.159        &  4.816  \\
           &    &   &   aug-cc-pVTZ   &   1  &  4.901    & 1.025  &   0.561    &   0.179    &   4.732 \\
          nitrous acid  &  $^{1}$A$'$      &  $(n)^{2} \to (\pi^{*})^{2} $    & 6-31+G*     &  1  &  4.302   & 0.745   &  0.330    &   0.005     & 8.170   \\
             &   &   &  aug-cc-pVDZ  &     1 &   4.680  &    1.117 &   0.443     &  0.112      &  8.057   \\
             &    &   &   aug-cc-pVTZ       &  1   &  5.320    &   1.136  &   0.555   &   0.096     &  7.969 \\
           formaldehyde   & $^{1}$A$_{1}$ &  $(n)^{2} \to (\pi^{*})^{2}$ &  6-31+G*  & 1  &   3.816  &   0.633  &  0.243   &  -0.073     &     10.859  \\
            &    &    & aug-cc-pVDZ  &   1  &  4.045  &   0.797   &  0.362    &  0.186    &     10.422  \\
            &    &    & aug-cc-pVTZ   &   1  &  4.483   &  0.774    &  0.359   & 0.117    &   10.426   \\
           cyclobutadiene &  $^{1}$A$_{g}$ &   $(\pi)^{2} \to (\pi^{*})^{2}$ & 6-31+G* &    1   & 3.002 & 0.585 & 0.238 & 0.209  & 4.073 \\
            &   &    & aug-cc-pVDZ &    1   & 3.246 & 0.665 & 0.281 & 0.227  & 4.046 \\
            &     &   &   aug-cc-pVTZ  &   1   & 3.895 & 0.741 & 0.393 & 0.273  & 4.036  \\
           cyclopentadienethione $\,$ &  $^{1}$A$_{1}$ &   $(n)^{2} \to (\pi^{*})^{2}$ & 6-31+G* &    1   & 4.890 & 0.914 & 0.596 &  0.043 & 5.780 \\
            &   &    &  aug-cc-pVDZ  &   1   & 5.108 & 1.012 & 0.703 &  0.095 & 5.615 \\
           glyoxal & $^{1}$A$_{g}$    &  $(n)^{2} \to (\pi^{*})^{2}$ & 6-31+G*   &    2  &  5.805  & 1.107   &   0.615    &  0.163   &   5.628 \\
     &   &   &  aug-cc-pVDZ       &   2 & 6.062   &  1.193  &  0.709   & 0.176  & 5.522 \\
              &  &   & aug-cc-pVTZ       &  2  &  6.687   &   1.271   &   0.861  & 0.163   & 5.492 \\
           pyrazine & $^{1}$A$_{g}$    &  $(n)^{2} \to (\pi^{*})^{2}$ & 6-31+G*   &   2  & 5.697    &   1.225 &  0.745  &  0.042    & 8.049 \\
            &    &    &  aug-cc-pVDZ   &   2  &    5.887      &  1.186  & 0.711  & -0.017   &  7.986 \\
           oxalyl fluoride &  $^{1}$A$_{g}$    & $(n)^{2} \to (\pi^{*})^{2}$  & 6-31+G*   &  2  &  6.651         &   1.201  &  0.805     &   0.218      &  9.056  \\
            &   &  &  aug-cc-pVDZ   &  2 &  6.895      &   1.276  &   0.898  &    0.202    &  8.976  \\
         \midrule
        \multicolumn{10}{c}{\textbf{1-CSF statistics}} \\
        \multicolumn{5}{c}{MSE $\pm$ Std. Dev.} & $4.06 \pm 0.83$ &  $0.80 \pm 0.19$     &  $0.38 \pm 0.15$   &  $0.14 \pm 0.10$   &         \\ 
        \multicolumn{5}{c}{MUE $\pm$ Std. Dev.} & $4.06 \pm 0.83$  &  $0.80 \pm 0.19$      &  $0.38 \pm 0.15$   &  $0.15 \pm 0.09$   &         \\ 
         \multicolumn{5}{c}{max error} & $5.32$ &  $1.14$     &  $0.70$   &  $0.30$   &         \\ 
          \multicolumn{10}{c}{\textbf{2-CSF statistics}} \\
        \multicolumn{5}{c}{MSE $\pm$ Std. Dev.} & $ 6.24 \pm 0.49 $ &  $ 1.21 \pm 0.06$     &  $0.76 \pm 0.10$   &  $0.14 \pm 0.09$   &         \\ 
        \multicolumn{5}{c}{MUE $\pm$ Std. Dev.} & $ 6.24 \pm 0.49 $ &  $1.21 \pm 0.06$     &  $0.76 \pm 0.10$   &  $0.14 \pm 0.08$   &         \\ 
         \multicolumn{5}{c}{max error} & $6.90$ &  $1.28$     &  $0.90$   &  $0.22$   &         \\ 
         \bottomrule         
         \end{tabular}
   
    \label{tab:ASCCtable}
\end{table*}



\begin{figure*}[ht]
    \centering
    \includegraphics[width=1.04\linewidth]{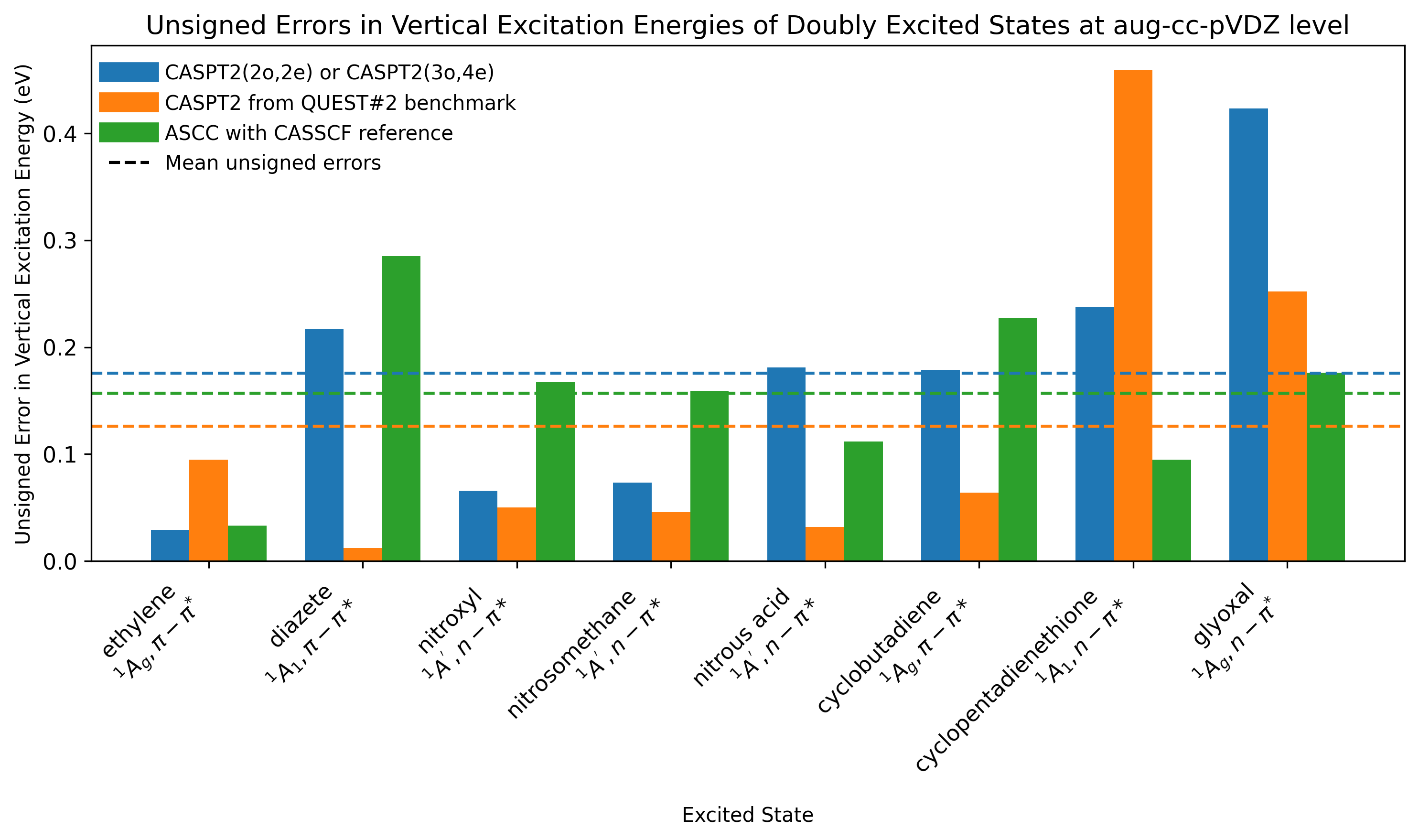}
    \caption{Comparison of CASPT2 and ASCC on seven 1-CSF states and one
             2-CSF state, with mean unsigned errors across these eight
             states shown as dashed lines.
             We report CASPT2 results both for the the same (2o,2e)
             or (3o,4e) active spaces used by ASCC and for the
             larger active spaces employed by the QUEST benchmark's
             CASPT2 results.
             }
    \label{fig:CASPT2}
\end{figure*}

In practice, multi-configurational double excitations are
often treated by CASPT2.
\cite{10.1063/1.462209,anderssonSecondorderPerturbationTheory1990,doi:https://doi.org/10.1002/9781119417774.ch10,https://doi.org/10.1002/qua.23052}
As both ASCC and CASPT2 start from a reference that correctly captures
the two-CSF character, we would expect both to perform relatively well,
and this is what we see for glyoxal in the data presented in Figure
\ref{fig:CASPT2}, where both methods produce errors smaller than
those of EOM-CCSDT.
For this state, as well as for a handful of single-CSF states,
we also compare whether limiting the active space of CASPT2
to the minimal ones employed by ASCC makes a significant difference.
As seen in the figure, CASPT2 has a slightly worse average performance
than ASCC when using its same minimal active space, but it
does slightly better on average if it uses larger active spaces.


\color{black}
We would like to stress that the success of ASCC in these
two-hole-one-particle 2-CSF states does not immediately answer
the question of how well it will perform in the wider realm
of other multi-configurational double excitations.
These initial 2-CSF results make clear that the theory can
reach beyond the 1-CSF regime, but further investigation will
be needed to more fully explore its multi-reference
capabilities.
For example, in cases with many
primary hole and many primary particle orbitals,
it is not yet clear  whether the mixed amplitudes will need
to include all levels of excitation within the primary
orbitals or if, as we do for the non-primary excitations,
the excitation level in the primary space can be safely capped.
If not, then the theory would display exponential scaling
in the number of primary orbitals and polynomial scaling
in the non-primaries, which would be manageable in some
applications but not in others.
Although these questions are clearly important, we do not attempt to
resolve them in this study, because there are a number of different
ways that a theory employing a de-excitation exponential to manipulate
the Aufbau coefficient could go about building a more
general multi-reference structure, and that question deserves
a study in its own right.
For now, we take these preliminary two-hole-one-particle 2-CSF results
as strong motivation for pursuing
more multi-reference generalizations
in the future.

\section{Conclusion}

We have explored a generalization of Aufbau suppressed coupled cluster
to doubly excited states.
Somewhat surprisingly, a zeroth order wave function initialization quite
similar to that used for singly excited states proves to be effective,
and the resulting first order amplitudes lead to the same asymptotic cost scaling, which matches that of the ground state CCSD. Moreover, working equations are exactly the same as in singly excited states case. Tests in single-CSF doubly excited states and a preliminary handful
of multi-CSF states reveal that even at the singles and
doubles level, the approach is already more accurate than EOM-CCSDT.
Indeed, the theory shows an accuracy in doubly excited
states that is on par with its previous strong performance in
singly excited states.

Looking forward, a key question will be how best to further generalize
the theory into a wider variety of multi-configurational excited states
that contain significant doubly excited character. One question worth further study is whether more
black-box starting points than CASSCF are possible.
For example, it may be worth testing the efficacy of replacing CASSCF
with a CC3/CC2 multi-level CC treatment \cite{myhre2014multi}
that performs CC3 within a full-valence active space,
from which natural orbitals and configuration coefficients could
be extracted to initialize a more accurate ASCC calculation.
At present, it is also not clear whether it will be necessary to include
all excitation levels within the primary orbitals (those actively
involved in the state's dominant configuration state functions)
or whether these excitations, like the external excitations, can be
safely capped at a low level.
Even if it is, the method should retain its asymptotic $N^6$ scaling
so long as the number of such orbitals does not grow with system size,
which at the very least should open applications in photochemistry in
which double excitations amongst a modest number of lone pair and
$\pi$-system orbitals currently frustrate an otherwise
coupled-cluster-friendly situation.
More broadly, the preliminary success in multi-configurational double
excitations raises the question of what other multi-configurational
states may be amenable to similar generalizations of
the Aufbau suppression approach.

\section*{Acknowledgments}
This work was supported by the National Science Foundation,
Award Number 2320936. Calculations were performed using
the Savio computational cluster resource provided by the
Berkeley Research Computing program at the University of
California, Berkeley. Q.J. and H.T. acknowledge that this material is based upon work supported by the National Science Foundation Graduate
Research Fellowship Program under Grant No. DGE 2146752.
Any opinions, findings, conclusions, or recommendations
expressed in this material are those of the authors and do not
necessarily reflect the views of the National Science
Foundation.

\section*{Data Availability Statement}

The data that support the findings of this study are available within the article and its supporting information. 

\section*{Supporting Information}
Tabulated vertical double excitation energies (in eV) computed with CASPT2, $\Delta$CCSD, IS-CCSD and ASCC in the aug-cc-pVDZ basis

\bibliography{main}

\section*{Supporting Information}
\renewcommand{\thesection}{S\arabic{section}}
\renewcommand{\theequation}{S\arabic{equation}}
\renewcommand{\thefigure}{S\arabic{figure}}
\renewcommand{\thetable}{S\arabic{table}}
\setcounter{section}{0}
\setcounter{figure}{0}
\setcounter{equation}{0}
\setcounter{table}{0}

\section{CASPT2 and ASCCSD}
\begin{table*}[ht]
    \centering
     \caption{Vertical excitation energies (in eV) of doubly excited states calculated at aug-cc-pVDZ level with CASPT2(2o,2e) for 1-CSF states and CASPT2(3o,4e) for 2-CSF states, QUEST CASPT2 and CASSCF-based ASCC. QUEST theoretical best estimates at aug-cc-pVDZ level are provided for reference.} 
\scriptsize
\begin{tabular}{lllccccc}
\toprule
\textbf{molecule} & \textbf{state} & \textbf{transition} &    \textbf{CSFs} & \textbf{CASPT2(2o,2e)/CASPT2(3o,4e)}  &  \textbf{QUEST CASPT2}   &  \textbf{ASCCSD} & \textbf{TBE} \\
\midrule
\hline \\
      ethylene  & $^{1}$A$_{g}$ &    $(\pi)^{2} \to (\pi^{*})^{2}$ &   1   &  13.039 &  13.163 & 13.101 & 13.068 \\
       diazete & $^{1}$A$_{1}$  &   $(\pi)^{2} \to (\pi^{*})^{2}$ & 1   & 6.418 & 6.623 & 6.920 & 6.635  \\
       nitroxyl &   $^{1}$A$'$    &  $(n)^{2} \to (\pi^{*})^{2} $  & 1 & 4.463 & 4.447 & 4.564 & 4.397  \\
         nitrosomethane &  $^{1}$A$'$   & $(n)^{2} \to (\pi^{*})^{2} $ & 1 & 4.889 & 4.862 & 4.975 & 4.816 \\
          nitrous acid  &  $^{1}$A$'$      &  $(n)^{2} \to (\pi^{*})^{2} $    & 1 & 8.238 &  8.089& 8.169 &   8.057 \\
             cyclobutadiene &  $^{1}$A$_{g}$ &   $(\pi)^{2} \to (\pi^{*})^{2}$ &  1   &  3.867   &  3.982  &  4.273  &  4.046 \\
             cyclopentadienethione $\,$ &  $^{1}$A$_{1}$ &   $(n)^{2} \to (\pi^{*})^{2}$  &  1   &  5.852  &  5.156  &  5.710 & 5.615   \\
             glyoxal & $^{1}$A$_{g}$    &  $(n)^{2} \to (\pi^{*})^{2}$ &  2   &  5.099  & 5.270    &  5.689  & 5.522 \\
            \midrule
             \multicolumn{8}{c}{\textbf{Statistics}} \\
        \multicolumn{4}{c}{MSE $\pm$ Std. Dev.} &   $-0.04 \pm 0.21$  &  $-0.07 \pm 0.18$  &  $0.16 \pm 0.07$  &       \\ 
        \multicolumn{4}{c}{MUE $\pm$ Std. Dev.} &  $0.18 \pm  0.12$  &  $0.13 \pm 0.14$  &  $0.16 \pm 0.07$     \\ 
         \multicolumn{4}{c}{max absolute error} & $0.42$  &  $0.46$  &       $0.28$  &  \\ 
         \bottomrule         
         \end{tabular}
   
    \label{tab:CASPT2table}
\end{table*}

\newpage

\section{$\Delta$CCSD and ASCCSD}
 \begin{table*}[ht]
    \centering
    \captionsetup{labelfont={bf, color=black},textfont={color=black}}
     \color {black} \caption{ \color{black} Vertical excitation energies (in eV) of doubly excited states calculated at aug-cc-pVDZ level with HF-$\Delta$CCSD, oo-$\Delta$CCSD and CASSCF-based ASCC. QUEST theoretical best estimates at aug-cc-pVDZ level are provided for reference.} 
\begin{tabular}{lllccccc}
\toprule
\textbf{molecule} & \textbf{state} & \textbf{transition} &    \textbf{CSFs} & \textbf{HF-$\Delta$CCSD}  &  \textbf{oo-$\Delta$CCSD}   &  \textbf{ASCCSD} & \textbf{TBE} \\
\midrule
\hline \\
      ethylene  & $^{1}$A$_{g}$ &    $(\pi)^{2} \to (\pi^{*})^{2}$ &   1   &  13.10 &  13.31 & 13.10 & 13.07 \\
       nitroxyl &   $^{1}$A$'$    &  $(n)^{2} \to (\pi^{*})^{2} $  & 1 & 4.65  & 4.66 & 4.56 & 4.40  \\
         nitrosomethane &  $^{1}$A$'$   & $(n)^{2} \to (\pi^{*})^{2} $ & 1 & 5.06 & 5.07 & 4.98 & 4.82 \\
          formaldehyde   & $^{1}$A$_{1}$ &  $(n)^{2} \to (\pi^{*})^{2}$ & 1 &  $-$  & 10.53  & 10.61 & 10.42 \\
             glyoxal & $^{1}$A$_{g}$    &  $(n)^{2} \to (\pi^{*})^{2}$ &  2   & 6.67  & 6.55    &  5.69  & 5.52 \\
              oxalyl fluoride & $^{1}$A$_{g}$    &  $(n)^{2} \to (\pi^{*})^{2}$ &  2   & 10.30  & 10.16    &  9.18  & 8.98 \\
            \midrule
             \multicolumn{8}{c}{\textbf{Statistics}} \\
        \multicolumn{4}{c}{MSE $\pm$ Std. Dev.} &   $0.60 \pm 0.53$  &  $0.51 \pm 0.43$  &  $0.15 \pm 0.06$  &       \\ 
         \multicolumn{4}{c}{MUE $\pm$ Std. Dev.} &   $0.60 \pm 0.53$  &  $0.51 \pm 0.43$  &  $0.15 \pm 0.06$  &       \\
         \multicolumn{4}{c}{max absolute error} & $1.32$  &  $1.18$  &       $0.20$  &  \\ 
         \bottomrule         
         \end{tabular}
   
    \label{tab:CASPT2table}
\end{table*}

\newpage

\section{IS-CCSD and ASCCSD}
 \begin{table*}[ht]
    \centering
    \captionsetup{labelfont={bf, color=black},textfont={color=black}}
     \color {black} \caption{ Vertical excitation energies (in eV) of doubly excited states calculated at aug-cc-pVDZ level with intermediate state EOM-CC approaches and CASSCF-based ASCC. QUEST theoretical best estimates at aug-cc-pVDZ level are provided for reference.} 
\small
\begin{tabular}{lllcccccc}
\toprule
\textbf{molecule} & \textbf{state} & \textbf{transition} &    \textbf{CSFs} & \textbf{IS-CCSD}  &  \textbf{IS-TCCSD}  & \textbf{IS-TCCSD}   &  \textbf{ASCCSD} & \textbf{TBE} \\  
       &  &  &     &   &  (M$_{\textrm
       {s}}=0$ triplet)  & (open-shell singlet)   &   &  \\
\midrule
\hline \\
      ethylene  & $^{1}$A$_{g}$ &    $(\pi)^{2} \to (\pi^{*})^{2}$ &   1   &  13.40 &  13.55 & 13.09   & 13.10 & 13.07 \\
       nitroxyl &   $^{1}$A$'$    &  $(n)^{2} \to (\pi^{*})^{2} $  & 1 &  4.39  & 4.39  & 4.31  & 4.56 & 4.40  \\
         nitrosomethane &  $^{1}$A$'$   & $(n)^{2} \to (\pi^{*})^{2} $ & 1 & 4.83   & 4.82  &  4.86 & 4.98 & 4.82 \\
          formaldehyde   & $^{1}$A$_{1}$ &  $(n)^{2} \to (\pi^{*})^{2}$ & 1 &  10.31  & 10.45    & 10.30   & 10.61 & 10.42 \\
             glyoxal & $^{1}$A$_{g}$    &  $(n)^{2} \to (\pi^{*})^{2}$ &  2   &  4.39   & 4.39   & 4.41    &  5.69  & 5.52 \\
             \midrule
             \multicolumn{9}{c}{\textbf{Statistics}} \\
        \multicolumn{4}{c}{MSE $\pm$ Std. Dev.} &   $-0.18 \pm 0.55$  &  $ -0.12 \pm 0.60$  &  $ -0.23   \pm  0.49  $  & $0.14 \pm 0.06$  &       \\ 
         \multicolumn{4}{c}{MUE $\pm$ Std. Dev.} &   $0.32 \pm  0.47 $   &  $0.33 \pm 0.49$  &  $ 0.37 \pm 0.52   $    & $0.14 \pm 0.06$  &       \\
         \multicolumn{4}{c}{max absolute error} & $ 1.12 $  &  $1.12$  &  $1.28$  &     $0.20$  &  \\ 
         \bottomrule         
         \end{tabular}
   
    \label{tab:IS-CCSD}
\end{table*}

 \newpage



\end{document}